\documentclass{emulateapj}
\slugcomment{}
\begin{document}
\title{A Tight $L_{\rm iso}-E_{\rm p, z}-\Gamma_0$ Correlation of Gamma-Ray Bursts}
\author{En-Wei Liang\altaffilmark{1,2,3}, Ting-Ting Lin\altaffilmark{1,2}, Jing L\"{u}\altaffilmark{1,2}, Rui-Jing Lu\altaffilmark{1,2}, Jin Zhang\altaffilmark{3,2}, Bing Zhang\altaffilmark{4,1,2}
}
\altaffiltext{1}{GXU-NAOC Center for Astrophysics and Space Sciences, Department of Physics, Guangxi University, Nanning 530004,
China; lew@gxu.edu.cn}
\altaffiltext{2}{Guangxi Key Laboratory for Relativistic Astrophysics, Nanning, Guangxi 530004, China}
\altaffiltext{3}{National Astronomical Observatories, Chinese Academy of Sciences, Beijing 100012, China}
\altaffiltext{4}{Department of Physics and Astronomy, University of Nevada, Las Vegas, NV 89154; zhang@physics.unlv.edu}
\begin{abstract}
We select a sample of 34 gamma-ray bursts (GRBs) whose $\Gamma_0$ values are derived with the onset peaks observed in the afterglow lightcurves (except for GRB 060218 whose $\Gamma_0$ is estimated with its radio data), and investigate the correlations among $\Gamma_0$,
the isotropic peak luminosity ($L_{\rm iso}$), and the peak energy ($E_{\rm p,z}$) of the $\nu f_\nu$ spectrum in the cosmological rest frame.
An analysis of pair correlations among these observables well confirms the results reported by the previous papers. More interestingly, a tight correlation among $L_{\rm iso}$, $E_{\rm p,z}$, and $\Gamma_0$ is found from a multiple regression analysis, which takes the form of $L_{\rm iso} \propto E_{\rm p,z}^{1.34\pm 0.14} \Gamma_0^{1.32\pm 0.19}$ or $E_{\rm p,z} \propto L_{\rm iso}^{0.55\pm 0.06}\Gamma_0^{-0.50\pm 0.17}$. Nine other GRBs whose $\Gamma_0$ are derived via the pair production opacity constraint also follow such a correlation. Excluding GRB 060218, the $L_{\rm iso}-E_{\rm p,z}-\Gamma_0$ correlation is valid, and it even holds in the jet co-moving frame. However, GRB 060218 deviates the $L^{'}_{\rm iso}-E^{'}_{\rm p}$ relation of typical GRBs in the jet co-moving frame with $3\sigma$. We argue that the $L_{\rm iso} - E_{\rm p, z} - \Gamma_0$ correlation may be more physical than the $L_{\rm iso} - E_{\rm p,z}$ correlation, since physically the relationship between the observed $L_{\rm iso}$ and $E_{\rm p,z}$ not only depends on radiation physics, but also depends on the bulk motion of the jet. We explore the possible origins of this correlation and discuss its physical implications for understanding GRB jet composition and radiation mechanism.
\end{abstract}
\section{Introduction} \label{sec:Introduction}

Gamma-ray bursts (GRBs) are the most luminous electromagnetic explosions in the Universe. The observed GRB spectra are typically fit by an empirical smoothly-jointed broken power-law function, the so-called Band function, characterized by a peak energy ($E_{\rm p}$) in the $\nu f_\nu$ spectrum, which ranges from keVs to MeVs (Band et al. 1993; Zhang et al. 2011). The physical origin of the $E_{\rm p}$ and the Band function spectrum is still subject to debate (Kumar \& Zhang 2015 for a recent review). One possibility is synchrotron radiation in internal shocks (e.g., M\'esz\'aros et al. 1994; Daigne \& Mochkovitch 1998; Daigne et al. 2011) or in internal magnetic dissipation regions (e.g., Zhang \& Yan 2011). A recent development of this model suggests that if the emission region is far from the central engine, the predicted spectra mimic the observed Band spectra (Uhm \& Zhang 2014), and can well fit the data with comparable confidence level as the Band function (Zhang et al. 2015). Another scenario attributes the $E_{\rm p}$ and Band function to emission from a dissipative photosphere of the fireball (e.g., Rees \& M\'esz\'aros 2005; Beloborodov 2010; Lazzati \& Begelman 2010; Lundman et al. 2013). These models invoke different compositions of GRB jets (e.g., matter dominated fireballs vs. Poynting-flux dominated jets). Differentiating them from the data has profound implications in understanding the physical mechanisms of GRBs.

Several important empirical correlations have been discovered with the GRB data. These correlations not only give important clues to understand GRB physics (even though some correlations still lack straightforward theoretical interpretations), but also some of them have been used to constrain cosmological parameters (e.g., Wang et al. 2015 for a recent review). For example, Amati et al. (2002) found a tight correlation between the isotropic gamma-ray energy $E_{\rm iso}$ and the cosmological rest-frame peak energy $E_{\rm p,z} = E_{\rm p} (1+z)$. A similar correlation was found between the isotropic luminosity and $E_{\rm p,z}$, both among different bursts (Yonetoku et al. 2004) and within a same burst (Liang et al. 2004). Ghirlanda et al. (2004) found a tighter relation by replacing $E_{\rm iso}$ with the geometrically-corrected energy of the GRB jets. The initial Lorentz factor ($\Gamma_0$) is a crucial parameter to understand GRB physics. Liang et al. (2010) discovered a correlation between ($\Gamma_0$) and $E_{\rm iso}$ (see also Ghirlanda et al. 2012). L\"{u} et al. (2012) showed that the isotropic luminosity ($L_{\rm iso}$) is also correlated with $\Gamma_0$.

Theoretically, the predicted $E_{\rm p,z}$ not only depends on luminosity $L_{\rm iso}$, but also depends on the bulk Lorentz factor $\Gamma_0$ of the outflow (see e.g., Table 1 of Zhang \& M\'esz\'aros 2002 for a summary). Therefore, it is interesting to search for possible multi-variable correlation among $E_{\rm p,z}$, $L_{\rm iso}$ (or $E_{\rm iso}$) and $\Gamma_0$. Some multi-variable correlations have been found for GRBs (e.g., Liang \& Zhang 2005; Rossi et al. 2008), which are useful to understand GRB physics.

This paper presents a multiple-variable regression analysis among $E_{\rm iso}$ (or $L_{\rm iso}$), $E_{\rm p,z}$, and $\Gamma_0$ for long duration GRBs. We compile a sample of long GRBs whose $E_{\rm iso}$ (or $L_{\rm iso}$), $E_{\rm p,z}$, and $\Gamma_0$ can be derived from the observation data (Section 2). Several correlations among these quantities are presented in Section 3. Physical implications are discussed in Section 4, and conclusions are drawn in Section 5. Notation $Q_n=Q/10^{n}$ is adopted in cgs units.

\section{Data}
We compile a sample of GRBs whose $E_{\rm iso}$, $L_{\rm iso}$, $E_{\rm p,z}$, and $\Gamma_0$ are available in the literature or can be calculated with observational data. $E_{\rm p,z}$ can be obtained by the measured $E_{\rm p}$ and redshift $z$. Both $E_{\rm iso}$ and $L_{\rm iso}$ are corrected to an energy band of $1-10^{4}$ keV in the burst rest frame.

The Lorentz factor $\Gamma_0$ is a key parameter in this analysis. Three methods have been proposed to estimate $\Gamma_0$ of a GRB fireball.
The first one is to use a smooth onset bump observed in optical afterglow lightcurves. By interpreting this bump as a result of deceleration of the fireball by an ambient medium in the thin shell regime (which is usually satisfied for a constant density medium), one can estimate $\Gamma_0$ with the peak time of the bump (M\'{e}sz\'{a}ros \& Rees 1993; Sari \& Piran 1999; Zhang et al. 2003; Molinari et al. 2007). The second method is based on the ``compactness" argument by interpreting the high energy cutoff of the prompt gamma-ray spectrum as pair production signature (e.g., Baring \& Harding 1997; Lithwick \& Sari 2001; Gupta \& Zhang 2008). The third approach is to use a blackbody component detected in the spectra of some GRBs (e.g., Pe'er et al. 2007; Peng et al. 2014; Zou et al. 2015). This method is based on the assumption of a matter-dominated fireball, and therefore is not reliable if the central engine carries a significant fraction of Poynting flux (Gao \& Zhang 2015).

The first method gives the most robust estimates to $\Gamma_0$, since the deceleration time weaklier depends on other model parameters than $\Gamma_0$. Such afterglow onset feature is observed in about 1/3 of GRBs with early optical afterglow observations (Li et al. 2012). Although the ``onset afterglow'' feature in some X-ray and GeV lightcurves is also observed (Xue et al. 2009; Ghisellini et al. 2010), early X-ray emission and GeV emission may have contamination from internal emission components. Thus, we only include GRBs with an onset bump detected in the optical band (most bursts from Liang et al. 2010, 2013 and references therein). We calculate their $\Gamma_0$ values with the afterglow onset peaks assuming that the fireballs of these GRBs are in an inter stellar medium with density profile $n=1$ cm$^{-3}$ and their radiation efficiencies are $\eta=0.2$ (e.g., Liang et al. 2013). The only exception is the nearby low luminosity GRB 060218 whose $\Gamma_0$ is robustly estimated from the radio data (Soderberg et al. 2006). The inclusion of this GRB stretches the dynamical range of the correlations significantly. $L_{\rm iso}$ values are measured in the 1s peak time of their lightcurves, and $E_{\rm iso}$ values are calculated with their gamma-ray fluences. Both $L_{\rm iso}$ and $E_{\rm iso}$ are corrected to $1-10^4$ keV band in the burst frame. Their $E_{\rm p}$ values are derived from the fits to their time-integrated spectra with the Band function\footnote{The $E_p$ of time-integrated spectrum is usually consistent with that derived from the spectrum around the peak time (Lyu et al. 2015).}. Altogether our sample includes 34 GRBs, which are reported in Table 1.

\section{Correlations among $E_{\rm iso}$, $L_{\rm iso}$, $E_{\rm p,z}$, and $\Gamma_0$}

With the available data, we conduct both single and multiple variable regression analysis to look for correlations among the parameters $E_{\rm iso}$, $L_{\rm iso}$, $E_{\rm p,z}$, and $\Gamma_0$. Notice that the results depend on the specification of dependent and independent variables (Isobe et al. 1990). Therefore, one may find discrepancy of the dependences among variables by specifying different dependent variables for a given data set, especially when the data have large error bars or large scatter. To avoid specifying independent and dependent variables in the best linear fits, in principle the algorithm of the bisector of two ordinary least-squares may be adopted. However, in connecting physical models, often some parameters (e.g., $y$) are believed to depend on other parameters (e.g., $x$), see more discussion in Section 4. We therefore use the Spearman correlation analysis to search for pair correlations among these parameters, and adopt the stepwise regression analysis method to perform a multiple regression analysis for multiple parameters by specifying a given $y$. We measure the dispersion ($\sigma$) of a regression model with standard deviation of $y^{\rm r}$ from $y$, where $\rm r$ marks the $y$ value derived from the regression model.

Figure 1 shows the pair correlations among $E_{\rm iso}$, $E_{\rm p,z}$, and $\Gamma_0$, or among $L_{\rm iso}$, $E_{\rm p,z}$, and $\Gamma_0$. They are derived in the logarithmic space, and the results are reported in Table 2. The regression lines together with their 2$\sigma$ dispersion regions are also shown in Figure 1. Correlations among $E_{\rm iso}$, $L_{\rm iso}$, $E_{\rm p,z}$, and $\Gamma_0$ are found, with a Spearman correlation coefficient $p>0.6$ and chance probability $p<10^{-4}$. Tight $E_{\rm iso}-E_{\rm p}$, $L_{\rm iso}-E_{\rm p}$, $E_{\rm iso}-\Gamma_0$, $L_{\rm  p, iso}-\Gamma_0$ correlations found in previous papers (e.g., Amati et al. 2002; Liang et al. 2004; Yonetoku et al. 2004; Liang et al. 2010; Lu et al. 2012; L\"u et al. 2012)\footnote{Ghirlanda et al. (2012) used a different method to estimate the $\Gamma_0$ values. Their method applies the Blandford-Mckee (1976) self-similar deceleration solution and extrapolates backwards to derive $\Gamma_0$.} are well confirmed. Combining the $E_{\rm iso}-E_{\rm p,z}$ (or $L_{\rm iso}-E_{\rm p,z}$) and the $E_{\rm iso}-\Gamma_0$ (or $L_{\rm iso}-\Gamma_0$) correlations, one may suspect a correlation between $E_{\rm p,z}$ and $\Gamma_0$. Such a correlation is indeed found in our sample, which is $E_{\rm p,z}\propto \Gamma_0^{0.77\pm 0.40}$ with a correlation coefficient of 0.63.

Next, we explore the possible correlations among $E_{\rm iso}$, $E_{\rm p,z}$, and $\Gamma_0$, or among $L_{\rm iso}$, $E_{\rm p,z}$, and $\Gamma_0$,
using the stepwise regression analysis method. Our model is $y^{\rm r}(x_1, x_2)=a+bx_1+cx_2$, where $y$, $x_1$, and $x_2$ stand for the observables (logarithmic). We measure the tightness of $y^{\rm r}(x_1, x_2)$ with the dispersion and linear coefficient of the pair correlation between $y^{\rm r}(x_1, x_2)$ and $y$. Our results are also reported in Table 2. Interestingly, much tighter correlations are found in our multiple variable regression analysis from the variable set \{$L_{\rm iso}$, $E_{\rm p,z}$, $\Gamma_0$\} than the pair correlations involving $L_{\rm iso}$. The dispersions of the $L_{\rm iso}-E_{\rm p,z}$ and $L_{\rm iso}-\Gamma_0$ pair correlations are 0.56 and 0.67, and their linear coefficients are 0.89 and 0.84, respectively. The dispersion of the relation $L^{\rm r}_{\rm p, iso}(E_{\rm p,z}, \Gamma_0)$ is reduced to 0.33 and the linear coefficient increases to 0.96. Figure 2 shows $y^{\rm r}(x_1, x_2)$ as a function of $y$ for the multi-variable correlations derived from the variable set \{$L_{\rm iso}$, $E_{\rm p,z}$, $\Gamma_0$\}.

The multi-variable correlations derived from the variable set \{$E_{\rm iso}$, $E_{\rm p,z}$, $\Gamma_0$\} are less significantly improved over the pair correlations involving $E_{\rm iso}$. The dispersions of the $E_{\rm iso}-E_{\rm p,z}$ and $E_{\rm iso}-\Gamma_0$ pair correlations are 0.57 and 0.63, and their linear coefficients are 0.83 and 0.77, respectively. The dispersion of the $E^{\rm r}_{\rm iso}(E_{\rm p,z}, \Gamma_0)$ relation is reduced to 0.41 and the linear coefficient is slightly increased to 0.88. We also show $E^{\rm r}_{\rm iso}(E_{\rm p,z}, \Gamma_0)$ as a function of $E_{\rm iso}$ in Figure 2. However, the dependence between $E_{\rm p,z}$ and $\Gamma_0$ in the relations of $E_{\rm p, z}^{\rm r}(E_{\rm iso}, \Gamma_0)$ and $\Gamma_0(E_{\rm iso}, E_{\rm p,z})$ is negligible. This could be due to the large dispersion of the $E_{\rm p,z}-\Gamma_0$ dependence in the current sample.

Our analysis shows that the dispersions of the $L_{\rm iso}-E_{\rm p,z}$ and $E_{\rm iso}-E_{\rm p,z}$ pair correlations are comparable. Nava et al. (2012) reported that the $L_{\rm iso}-E_{\rm p,z}$ correlation has slightly larger scatter than the $E_{\rm iso}-E_{\rm p,z}$ correlation. This  would be due to the variation of small sub-samples used for analysis. Note that $E_{\rm iso}$ is a time-integrated quantity. $E_{\rm p}$ usually evolves significantly within a GRB and even within  a GRB pulse (e.g., Lu et al. 2012). The $E_p$ of time-integrated spectrum is usually consistent with that derived from the spectrum around the peak time (Lyu et al. 2015). Therefore, the time-integrated effect of $E_{\rm iso}$ would result in significant scatter to the $E_{\rm iso}-E_{\rm p,z}$ correlation. It was suggested that the $L_{\rm iso}-E_{\rm p,z}$ correlation may be the basis of the time-integrated $E_{\rm iso}-E_{\rm p,z}$ correlation (Lu et al. 2012; Lyu et al. 2015)\footnote{Since GRB lightcurves tend to peak at different time in different energy
bands, and, moreover, the time on which $L_{\rm iso}$ is calculated is often not homogeneous in the cosmological rest-frame. This would make extra scatters to the $L_{\rm iso}-E_{\rm p,z}$ correlation.}. The 3-parameter correlations involving $\Gamma_0$ improve significantly when using $L_{\rm iso}$. This may be due to the fact that $L_{\rm iso}$  has a double dependence on $\Gamma_0$ (both photon energy and time), but $E_{\rm iso}$ only has a dependence on $\Gamma_0$ for photon energy.

We focus on the \{$L_{\rm iso}$, $E_{\rm p,z}$, $\Gamma_0$\} correlation in our following analysis. This correlation takes three forms, but two forms are of theoretical interest in the GRB physics (e.g., Zhang \& M\'{e}sz\'{a}ros 2002) and GRB cosmology (e.g., Dai et al. 2004).

First, in terms of $L_{\rm iso}$, it can be expressed as
\begin{equation}
L_{\rm iso, 52}=10^{-6.38\pm 0.35}{(E_{\rm p,z}/{\rm keV})}^{1.34\pm 0.14}\Gamma_{0}^{1.32\pm 0.19},
\end{equation}
with a correlation coefficient of 0.96 and a dispersion of $\delta= 0.33$ dex. By adding $\Gamma_0$, this relation is tighter than the $L_{\rm iso}-E_{\rm p,z}$ (Yonetoku) relation and $E_{\rm iso}-E_{\rm p,z}$ (Amati) relation. It suggests that the relatively large dispersion in the Yonetoku-relation and Amati-relation is likely due to the lack of consideration of the role of $\Gamma_0$, a key parameter for the relativistic outflows of GRBs.

The second format of the correlation is expressed in terms of $E_{\rm p,z}$, which reads
\begin{equation}\label{Ep_model}
E_{\rm p,z}=10^{3.71\pm 0.38} {\rm keV}\times L^{0.55\pm 0.06}_{\rm iso, 52}\Gamma^{-0.50\pm 0.17}_0,
\end{equation}
with a correlation coefficient of 0.92 and a dispersion of $\delta = 0.20$. This format more directly carries a physical meaning, which we discuss in the next section.

Our analysis is based on the GRBs whose $\Gamma_0$ are measured with the deceleration peak time in the afterglow lightcurves, except for GRB 060218. This method is the most robust one to estimate $\Gamma_0$. For a self-consistency check, we also introduce other GRBs whose $\Gamma_0$ are derived from other methods. For example, Tang et al. (2015) systematically searched for the high energy spectral cutoffs from the joint LAT/GBM observations of GRBs and estimated $\Gamma_0$ for 9 GRBs using the opacity argument. They are consistent with the derived $L_{\rm iso}(E_{\rm p,z}, \Gamma_0)$ correlation, as shown in Figure 2.

With the measured $\Gamma_0$ for the GRBs in our sample, we further investigate the possible correlations among $L^{'}_{\rm iso}$ (or $E^{'}_{\rm iso}$ ), $E_p^{'}$, and $\Gamma_0$ in the jet-comoving frame (e.g., Ghirlanda et al. 2013), where $L^{'}_{\rm iso}=L_{\rm iso}/\Gamma_0^2$, $E^{'}_{\rm iso}=E_{\rm iso}/\Gamma_0$, $E_p^{'}=E_{\rm p, z}/\Gamma_0$. Our regression analysis results are reported in Table 3. It is found that the pair correlations are much weaker than those among $L_{\rm iso}$ (or $E_{\rm iso}$), $E_{\rm p,z}$, and $\Gamma_0$ and their chance probabilities are larger than $10^{-4}$. These results suggest that the observed pair correlations are likely governed by the Doppler boosting effect. The three-parameter regressions show that the $L^{'}_{\rm iso}-E_p^{'}-\Gamma_0$ relation still exists, as shown in Figure 3. The difference between the $L^{'}_{\rm iso}-E_p^{'}-\Gamma_0$ and the $L_{\rm iso}-E_{\rm p, z}-\Gamma_0$ relations is the index of $\Gamma_0$. This is reasonable since $L^{'}_{\rm iso}$ and $E_p^{'}$ are corrected by the $\Gamma_0$ factor.

Note that GRB 060218 is peculiar with extremely low $L_{\rm iso}$, $E_{\rm p}$, and $\Gamma_0$ in comparison with the typical GRBs in our sample\footnote{It was suggested that this event may be of a distinct population different from the typical GRBs (e.g., Soderberg et al. 2006, Liang et al. 2007), but it follows the $E_{p,z}-E_{\rm iso}$ relation and the spectral lag-$L_{\rm iso}$ relation of typical GRBs (Amati et al. 2007; Liang et al. 2006).}. The inclusion of this event significantly expands the dynamical range of the correlations discussed above. Excluding this event, we find that the $L_{\rm iso}-E_{\rm p, z}-\Gamma_0$ relation still holds, showing as $L_{\rm iso, 52}=10^{-5.36\pm 0.49}{(E_{\rm p,z}/{\rm keV})}^{1.23\pm 0.14}\Gamma_{0}^{1.00\pm 0.21}$ and $E_{\rm p,z}=10^{3.57\pm 0.39} {\rm keV}\times L^{0.59\pm 0.07}_{\rm iso, 52}\Gamma^{-0.45\pm 0.17}_0$.  Within error bars, the indices are consistent with that derived from our GRB 060218-included sample. This indicates that the $L_{\rm iso}-E_{\rm p, z}-\Gamma_0$ relation is not from broad ranges being due to this peculiar event. We also make correlation analysis among $L^{'}_{\rm iso}$ (or $E^{'}_{\rm iso}$ ), $E_p^{'}$, and $\Gamma_0$ by excluding GRB 060218 from our sample. The results are reported in Table 4. One can find that the $L^{'}_{\rm iso}-E^{'}_{\rm p}-\Gamma_0$ relation still exists (see Figure 3), but the indices of $\Gamma_0$ are getting smaller and have a larger uncertainty than that in the $L_{\rm iso}-E_{\rm p, z}-\Gamma_0$ relation. It is also interesting that an $L^{'}_{\rm iso}-E^{'}_{\rm p}$ relation is found for typical GRBs, i.e., $\log L^{\rm '}_{\rm iso, 46}=(1.19\pm 0.08)+(1.18\pm0.13)\times \log (E^{\rm '}_{\rm p}/{\rm keV}$) with a Spearman linear correlation coefficient of 0.85 and chance probability $p<10^{-4}$, but GRB 060218 deviates from this relation at a 3 $\sigma$ confidence level (Figure 3). This may hint that the radiation physics of this event may be different from typical GRBs (e.g., Campana et al. 2006; Wang et al. 2007).

\section{Physical implications}

Different GRB prompt emission models have different predictions of the rest-frame peak energy $E_{\rm p,z}$. Zhang \& M\'esz\'aros (2002) summarized these predictions in their Table 1. In general, $E_{\rm p,z}$ is not only a function of the outflow luminosity $L$, which may be proportional to the observed luminosity $L_{\rm p,iso}$, but also depends on $\Gamma_0$. The $E_{\rm p,z}$ predictions of different models have different indices on $L$ and $\Gamma_0$. With the available correlation (Eq. \ref{Ep_model}), one can place strong constraints on various models.

An immediate inference from Table 1 of Zhang \& M\'esz\'aros (2002) is that the predicted $E_{\rm p,z}$ of the simplest baryonic photosphere models (called ``innermost model'' in the paper) and the external shock models all have wrong dependence on both $L$ and $\Gamma_0$. The external shock model is now essentially ruled out with the \emph{Swift} early X-ray afterglow data that show a steep decay phase (see Zhang et al. 2006 for a detailed discussion). For the photosphere model, one has $E_{\rm p,z} \propto T_0$ (central engine temperature) $\propto L^{1/4}$ if the photosphere radius $R_{\rm ph}$ is below the coasting radius $R_{\rm c}$. There is no $\Gamma_0$-dependence in this regime, which is not supported by the data. When $R_{\rm ph}>R_{\rm c}$, one has $E_{\rm p,z} \propto T_{ph} = T_0 (R_{\rm ph}/R_{\rm c})^{-2/3} \propto L_w^{1/4} (L_w/\Gamma_0^4)^{-2/3} \propto L_w^{-5/12} \Gamma_0^{8/3}$ (Zhang \& M\'esz\'aros 2002), where $L_w$ is the wind luminosity. This apparently conflicts with the observations. Noticing that the photosphere luminosity scales as $L \sim L_{ph} \sim L_w (R_{ph}/R_c)^{-2/3}$, one gets $L_w \propto L^3 \Gamma_0^{-8}$. This gives
\begin{equation}
E_{\rm p,z} \propto T_{ph} \propto L^{-5/4} \Gamma_0^6,
\label{eq:Ep-ph}
\end{equation}
which violates the observations badly\footnote{Fan et al. (2012) argued that the photosphere model can interpret four two-parameter correlations. However, the model cannot interpret the three-parameter correlation discovered in our paper. Their Eq.(3) is essentially $E_{\rm p,z} \propto L^{-1/4} \Gamma^2 Y_b^{1/2}$. Noticing $Y_b =L/L_w$, one again derives Eq.(\ref{eq:Ep-ph}).}. For dissipative photospheres, $E_{\rm p,z}$ may be different from the photosphere temperature estimated above. However, there is no straightforward physics to drive $E_{\rm p,z}$ to satisfy Eq.(\ref{Ep_model}). We therefore conclude that the data do not favor the photosphere models of GRB prompt emission.

A favorable model is synchrotron radiation in an internal emission region (internal shocks, M\'{e}sz\'{a}ros et al. 1994; or an internal magnetic dissipation site, Zhang \& Yan 2011). Within this model, one can generally write (Zhang \& M\'{e}sz\'{a}ros 2002)
\begin{equation}
E_{\rm p,z} \propto L^{1/2} R^{-1},
\label{Ep-syn}
\end{equation}
where $R$ is the radius of the emission region from the central engine. One can see that the 1/2 index for $L$ match Eq.(\ref{Ep_model}), which suggests that synchrotron radiation is likely the right emission mechanism. If the outflow is not magnetized, the emission radius is the internal shock radius, i.e. $R \sim \Gamma_0^2 c \delta t \propto \Gamma_0^2$. From Eq.(\ref{Ep-syn}) one gets $E_{\rm p,z} \propto L^{1/2} \Gamma_0^{-2}$. The index for $\Gamma_0$ is too steep (-2) compared with the data ($-0.50\pm 0.17$) (Eq.\ref{Ep_model}). This suggests that there must be another factor that moderates $R$ to have a shallower dependence on $\Gamma_0$. If the outflow is Poynting-flux-dominated, so that synchrotron emission comes from the ICMART (internal-collision-induced magnetic reconnection and turbulence) region, then the predicted $E_{\rm p,z}$ has an extra dependence on $\sigma$, i.e. (Eq. 58 of Zhang \& Yan 2011)
\begin{equation}
 E_{\rm p,z} \propto L^{1/2} R^{-1} \sigma^2.
\end{equation}
This extra dependence can make it possible to interpret the observed correlation. In particular, the data demand $R^{-1} \sigma^2 \propto \Gamma_0^{-0.50\pm 0.17}$. Zhang \& Yan (2011) argued that a larger $\sigma$ tends to increase $R$, since more collisions are required to finally reach the critical point for ICMART discharge, so that $R^{-1} \sigma^2$ should have a shallower $\Gamma_0$ dependence than $\Gamma_0^{-2}$. Even though more detailed numerical modeling is needed to reveal the nature of our correlation (\ref{Ep_model}), this qualitative picture seems to be consistent with the data.

\section{Conclusions and Discussion}
We presented a multiple linear regression analysis to key observables of the GRB outflows, i.e., $L_{\rm iso}$ ($E_{\rm iso}$), $E_{\rm p,z}$, and $\Gamma_0$. The analysis of pair correlations among these observables well confirms several previously reported correlations. Most interestingly, we find a new tight correlation among $L_{\rm iso}$, $E_{\rm p,z}$, and $\Gamma_0$ from our multiple regression analysis. We argue that this tight $L_{\rm iso}(E_{\rm p,z}, \Gamma_0)$ correlation is more physical than the $L_{\rm iso}-E_{\rm p,z}$ and $L_{\rm iso}-\Gamma_0$ correlations, and it may be directly related to both radiation physics and bulk motion of the outflows.

The tight $E_{\rm p,z}(L_{\rm iso}, \Gamma_0)$ correlation sheds light into the origin of GRB prompt emission. We show that the photosphere models have difficulties to account for the correlation, and synchrotron radiation is likely the radiation mechanism of GRB prompt emission. This is consistent with other independent arguments in favor of synchrotron radiation (Zhang et al. 2013; Wang et al. 2014; Uhm \& Zhang 2014; Zhang et al. 2015). Within the synchrotron model, the internal shock model predicts a too steep dependence on $\Gamma_0$, and hence, not favored. The ICMART magnetic dissipation model (Zhang \& Yan 2011) has the general merit to account for the correlation, even though no quantitative proof to the correlation is available.

\begin{acknowledgements}
We thank the anonymous referee for his/her valuable suggestions. This work is supported by the National Basic Research Program (973 Programme) of China (Grant No. 2014CB845800), the National Natural Science Foundation of China (Grant Nos. 11533003, 11573034, 11363002, 11373036), the Guangxi Science Foundation (2013GXNSFFA019001 and 2014GXNSFBA118009), the Strategic Priority Research Program ``The Emergence of Cosmological Structures''(Grant No. XDB09000000).
\end{acknowledgements}

\clearpage

\begin{deluxetable}{llllll}
\tabletypesize{\scriptsize}
\tablewidth{350pt}
\tablecaption{The redshift ($z$), isotropic luminosity ($L_{\rm iso}$) and energy $E_{\rm iso}$ as well as the peak energy of the $\nu f_\nu$ spectrum ($E_{\rm p}$) of prompt gamma-rays, and the initial Lorentz factor of GRB fireball ($\Gamma_0$) for the GRBs in our sample.}
\tablecolumns{6}
\tablehead{
\colhead{GRB}&
\colhead{$z$}&
\colhead{$L_{\rm iso}$ ($10^{52}$ erg/s)}&
\colhead{$E_{\rm iso}$($10^{52}$ erg)}&
\colhead{$E^{'}_{\rm p}$ (keV)}&
\colhead{$\Gamma_0$}
}
\startdata
990123	&  	1.6	    &$27.5\pm1.2$           &  $356\pm7$             & $1334_{-57}^{+50}$   & $600\pm 80$   \\ 
050820A	&  	2.615	&$3.22_{-0.45}^{+0.30}$ &$15.9_{-1.5}^{+1.1}$    &$889_{-239}^{+459}$   &$	282_{-14}^{+29}$  \\ 
050922C	&  	2.198	&$6.6_{-0.54}^{	+0.29}$ &$5.44_{-0.46}^{+0.24}$    &$629_{-118}^{+204}$   &	274  \\ 
060210	&  	3.91	&$7.40_{-1.32}^{+1.98}$ &$55.2_{-3.0}^{	+18.4}$    &$732_{-172}^{+1964}$   &$	 264\pm4$  \\
060218	&  0.0331	&$	4.27_{-1.66}^{+1.54}\times 10^{-6}$& $	93.2_{-7.6}^{+7.3}\times 10^{-5}$   & $5.1\pm0.3$   &	2.3 \\ 
060418	&  	1.489	&$1.9_{-0.15}^{+0.11}$  &$14.3\pm0.4$    & 572.5   &$263_{-7}^{+23}$ \\
060605	&  	3.78	&$0.95\pm0.15$          &$2.83\pm0.45$    &$490	\pm251$   &$	197	_{	-6	}^{+30}$  \\
060607A	&  	3.082	&$2.0\pm0.27$           &$9_{-2	}^{	+7	}$    &$	575	\pm200	$   &$	296	_{-8}^{+28}$  \\
060904B	&  	0.703	&$0.074\pm0.014$&$	0.364\pm0.074	$    &$	135\pm41	$   &$	108	\pm10	$  \\
061007	&  	1.261	&$14.3_{-1.8}^{+2.3}$&$	95.4_{	-4.6}^{	+6.5}$    & $902_{ -41}^{ +43}	$   &$	436	 \pm3	 $  \\
061121	&  	1.314	&$14.1\pm	0.15	$&$	26.1\pm3$    &$	1289\pm	153	$   &$	175\pm	2$  \\
070110	&  	2.352	&$0.451\pm0.075	$&$	5.5\pm	1.5$    &$	370	\pm	170	$   &$	127\pm	4$  \\
070208	&  	1.165	&	0.093&$0.295_{-1.35}^{+2.19}$    &$66_{-33}^{+179}$   &$115_{	-20	}^{	 +23	}$  \\
070419A	&  	0.97	&	0.0098&$1.74_{-1.12}^{+1.70}$    &$27_{-19}^{+16}$   &$91_{-3}^{+11}$  \\
071010B	&  	0.947	&$	0.555_{	-0.026	}^{	+0.022}$&$	2.6	_{-0.14}^{+0.19}$    & $101\pm13$   &$	 209\pm4$  \\
071112C	&  	0.822	&	1.047&	$61.66\pm36.91$ &  $422	_{-87}^{+137}$   &	244	  \\
080129	&  	4.394	&	2.69&	7	    &$	1349_{	-809}^{	+2643}$   &	65	 \\
080319C	&  	1.95	&$	9.5	\pm	0.12$&$	15\pm0.79$    &$	1752\pm505	$   &$	228	\pm	5	$  \\
080710	&  	0.845	&	0.079    &   $0.8_{-0.4}^{+0.8}$    &$	300	_{-200}^{+500}$   &$63_{-4}^{+8	}$  \\
080810	&  	3.35	&$	9.56\pm	0.83	$&$	40.5\pm	2.9	$    & $	1364\pm320$   &$	 409\pm	34$  \\
081008	&  	1.967	&$	0.55\pm	0.01$ &    $6.61_{-1.20}^{+1.62}$    &$267_{-62}^{+335}$   &	250  \\
081109A	&  	0.98	&$	0.20\pm0.03   $    &   $4.07_{-1.70}^{+1.91}$    &$208_{-50}^{+303}$   &	 68	 \\
081203A	&  	2.1	    &$	2.81\pm	0.19	$&$	35\pm3	$    &$	1541\pm757	$   &$	219_{-6}^{+21	}$  \\
090102	&  	1.547	&$	5.83_{-0.82}^{+0.84	}$&$	21.5\pm	2	$    & $	1149 _{-148}^{+186}$   &	61	\\
090424	&  	0.544	&$	2.07_{-0.12}^{+0.13	}$&$	4.35_{-0.15}^{+0.16	}$    & $273\pm 5$   &$	300	\pm	79	$  \\
090812	&  	2.452	&$	10_{-1.3}^{+0.9}$&$	42.1\pm	5.5	$    & $1975_{ -549}^{+867}$   &	501	 \\
091024	&  	1.092	&$	1.00\pm0.22$&$	28\pm3	$    &$	794\pm	231$   &	69 \\
091029	&  	2.752	&$	1.72\pm	0.10$&$	9.5	_{	-0.48	}^{	+0.86	}$    & $	230\pm66$   &	221	 \\
100621A	&  	0.542	&$	0.316\pm0.024$&$	4.37\pm	0.5$    &$	146\pm	23.1	$   &	52	\\
100728B	&  	2.106	&$	1.86\pm0.12	$&$	3\pm0.3	$    &$	404\pm29	$   &	373 \\
100906A	&  	1.727	&$	2.45\pm	0.09	$&$	33.4\pm	3	$    &$	158\pm	16$   &	369  \\
110205A	&  	2.22	&$	2.50	\pm0.34	$&$	56\pm6$    &$	715\pm	239$   &	177	 \\
110213A	&  	1.46	&$	2.09\pm	0.06	$&$	6.4\pm	0.6	$    &$	241\pm	$13	   &	223	\\
121217A	&   3.1	    &$	3.51\pm0.54$& 62    &$	754\pm230$   &	247
\enddata
\end{deluxetable}

\clearpage
\begin{deluxetable}{lllll}
\tabletypesize{\scriptsize}
\tablewidth{500pt}
\tablecaption{Results of Our Linear Regression Analysis for $L_{\rm iso}$, $E_{\rm p,z }$, and $\Gamma_0$ in the observer frame with all GRBs in our sample, in which $r$ is the Spearman correlation coefficient, $p$ is the chance probability, and $\delta$ is the dispersion of the reported relations.}
\tablecolumns{5}
\tablehead{
\colhead{Relations}&
\colhead{Expressions}&
\colhead{$r$}&
\colhead{$p$}&
\colhead{$\delta$}
}

\startdata
$L_{\rm iso}(E_{\rm iso})$ & $\log L_{\rm iso,52}=(-0.99\pm 0.13)+(1.12\pm0.10)\times \log E_{\rm iso,52}$ & 0.90 & $<10^{-4}$ & 0.54 \\
$L_{\rm iso}(E_{\rm p,z})$ & $\log L_{\rm iso, 52}=-(5.10\pm 0.47)+(1.97\pm0.18)\times \log (E_{\rm p,z}/{\rm keV}$) & 0.89 & $<10^{-4}$ & 0.56\\
$L_{\rm iso}(\Gamma_0)$ & $\log L_{\rm iso, 52}=-(5.35\pm0.63)+(2.42\pm0.28)\times \log \Gamma_0$ & 0.84 &$<10^{-4}$ & 0.67 \\
$E_{\rm iso}(E_{\rm p,z})$ & $\log E_{\rm iso, 52}=-(2.95\pm 0.48)+(1.49\pm0.18)\times \log (E_{\rm p,z}/{\rm keV})$ & 0.83 & $<10^{-4}$ & 0.57\\
$E_{\rm iso}(\Gamma_0)$ & $\log E_{\rm iso, 52}=-(3.06\pm0.59)+(1.78\pm0.26)\times \log \Gamma_0$ & 0.77 & $<10^{-4}$ & 0.63 \\
$E_{\rm p,z}(\Gamma_0)$ & $\log (E_{\rm p,z}/{\rm keV})=(0.77\pm0.40)+(0.82\pm0.18)\times \log \Gamma_0$ & 0.63 & $<10^{-4}$ & 0.43 \\
\hline
$L^{\rm r}_{\rm iso}(E_{\rm p,z},\Gamma_0)$ & $\log L^{\rm r}_{\rm iso, 52}=-(6.38\pm0.35)+(1.34\pm0.14)\times \log (E_{\rm p,z}/{\rm keV})+(1.32\pm0.19)\times \log\Gamma_0$ & 0.96 & $<10^{-4}$& 0.33 \\
$E_{\rm p,z}^{\rm r}(L_{\rm iso},\Gamma_0)$ & $\log (E_{\rm p,z}^{\rm r}/{\rm keV})=(3.71\pm0.38)+(0.55\pm0.06)\times \log L_{\rm iso, 52}-(0.50\pm0.17)\times \log \Gamma_0$ & 0.92 & $<10^{-4}$ &0.20  \\
$\Gamma^{\rm r}_0(L_{\rm iso},E_{\rm p,z})$ & $\log \Gamma^{\rm r}_0=(3.33\pm0.38)+(0.46\pm0.07)\times \log L_{\rm iso, 52}-(0.43\pm0.15)\times \log (E_{\rm p,z}/{\rm keV})$ & 0.88 & $<10^{-4}$ & 0.18 \\
$E^{\rm r}_{\rm iso}(E_{\rm p,z},\Gamma_0)$ & $\log E^{\rm r}_{\rm iso, 52}=-(3.81\pm0.47)+(0.98\pm0.19)\times \log (E_{\rm p,z}/{\rm keV})+(0.97\pm0.25)\times \log\Gamma_0$ & 0.88 & $<10^{-4}$ & 0.41 \\
$E_{\rm p,z}^{\rm r}(E_{\rm iso},\Gamma_0)$ & $\log (E_{\rm p,z}^{\rm r}/{\rm keV})=(2.17\pm0.41)+(0.46\pm0.09)\times \log E_{\rm iso, 52}+(0.00\pm0.21)\times \log \Gamma_0$ & 0.82 & $<10^{-4}$ & 0.26 \\
$\Gamma^{\rm r}_0(E_{\rm iso},E_{\rm p,z})$ & $\log \Gamma^{\rm r}_0=(1.92\pm0.34)+(0.33\pm0.09)\times \log E_{\rm iso, 52}+(0.00\pm 0.15)\times \log (E_{\rm p,z}$/{\rm keV}) & 0.77 & $<10^{-4}$& 0.21
\enddata
\end{deluxetable}

\begin{deluxetable}{lllll}
\tabletypesize{\scriptsize}
\tablewidth{500pt}
\tablecaption{Results of Our Linear Regression Analysis for $L^{'}_{\rm iso}$, $E^{'}_{\rm p}$, and $\Gamma_0$ in the jet co-moving frame with all GRBs in our sample, in which $r$ is the Spearman correlation coefficient, $p$ is the chance probability, and $\delta$ is the dispersion of the reported relations.}
\tablecolumns{5}
\tablehead{
\colhead{Relations}&
\colhead{Expressions}&
\colhead{$r$}&
\colhead{$p$}&
\colhead{$\delta^{*}$}
}
\startdata
$L^{\rm '}_{\rm iso}(E^{\rm '}_{\rm iso})$ & $\log L^{\rm '}_{\rm iso, 46}=(1.08\pm 0.11)+(0.73\pm0.11)\times \log E^{'}_{\rm iso,50}$ & 0.75 & $<10^{-4}$ & 0.46 \\
$L^{\rm '}_{\rm iso}(E^{\rm '}_{\rm p})$ & $\log L^{\rm '}_{\rm iso, 46}=-(0.31\pm 0.27)+(0.89\pm0.47)\times \log (E^{\rm '}_{\rm p}/{\rm keV}$) & 0.32 & 0.07  & --\\

$L^{\rm '}_{\rm iso}(\Gamma_0)$ & $\log L^{\rm '}_{\rm iso, 46}=(0.65\pm0.63)+(0.42\pm0.28)\times \log \Gamma_0$ & 0.26 & 0.14 & -- \\
$E^{\rm '}_{\rm iso}(E^{\rm '}_{\rm p})$ & $\log E^{\rm '}_{\rm iso, 50}=(0.38\pm 0.14)+(0.82\pm0.25)\times \log (E^{\rm '}_{\rm p}/{\rm keV})$ & 0.50 & $2.36\times 10^{-3}$ & --\\

$E^{\rm '}_{\rm iso}(\Gamma_0)$ & $\log E^{\rm '}_{\rm iso, 50}=-(1.06\pm0.60)+(0.78\pm0.26)\times \log \Gamma_0$ & 0.47 & $5.39\times10^{-3}$ & -- \\
$E^{\rm '}_{\rm p}(\Gamma_0)$ & $\log (E^{\rm '}_{\rm p}/{\rm keV})=(0.77\pm0.40)-(0.18\pm0.18)\times \log \Gamma_0$ & -0.17 & 0.33 & -- \\
\hline
$L^{\rm 'r}_{\rm iso}(E^{\rm '}_{\rm p},\Gamma_0)$ & $\log L^{\rm 'r}_{\rm iso, 46}=-(0.38\pm0.35)+(1.34\pm0.14)\times \log (E^{\rm '}_{\rm p}/{\rm keV})+(0.66\pm0.15)\times \log\Gamma_0$ & 0.87 & $<10^{-4}$& 0.30 \\
$E_{\rm p}^{\rm ' r}(L^{\rm '}_{\rm iso},\Gamma_0)$ & $\log (E_{\rm p}^{\rm ' r}/{\rm keV})=(0.42\pm0.22)+(0.55\pm0.06)\times \log L^{\rm '}_{\rm iso, 46}-(0.41\pm0.10)\times \log \Gamma_0$ & 0.86 & $<10^{-4}$ & 0.19 \\
$\Gamma^{\rm r}_0(L^{\rm '}_{\rm iso},E^{\rm '}_{\rm p})$ & $\log \Gamma^{\rm r}_0=(1.63\pm0.17)+(0.59\pm0.13)\times \log L^{\rm '}_{\rm iso, 46}-(0.89\pm0.21)\times \log (E^{\rm '}_{\rm p}/{\rm keV})$ & 0.64 & $<10^{-4}$ & 0.21 \\
$E^{\rm 'r}_{\rm iso}(E^{\rm '}_{\rm p},\Gamma_0)$ & $\log E^{\rm 'r}_{\rm iso, 50}=-(1.81\pm0.47)+(0.98\pm0.19)\times \log (E^{\rm '}_{\rm p}/{\rm keV})+(0.96\pm0.20)\times \log\Gamma_0$ & 0.76 & $<10^{-4}$ & 0.35 \\
$E_{\rm p}^{\rm ' r}(E_{\rm iso}^{\rm '},\Gamma_0)$ & $\log (E_{\rm p}^{\rm ' r}/{\rm keV})=(1.25\pm0.32)+(0.46\pm0.09)\times \log E_{\rm iso, 50}^{\rm '}+(0.54\pm0.15)\times \log \Gamma_0$ & 0.68 & $<10^{-4}$ & 0.22 \\
$\Gamma^{\rm r}_0(E^{\rm '}_{\rm iso},E^{\rm '}_{\rm p})$ & $\log \Gamma^{\rm r}_0=(2.12\pm0.08)+(0.44\pm0.09)\times \log E^{\rm '}_{\rm iso, 50}-(0.53\pm 0.15)\times \log (E^{\rm '}_{\rm p}$/{\rm keV}) & 0.66 & $<10^{-4}$& 0.21 \\
\enddata
\tablenotetext{*}{Reported only for the relation with $p<10^{-4}$.}

\end{deluxetable}

\begin{deluxetable}{lllll}
\tabletypesize{\scriptsize}
\tablewidth{500pt}
\tablecaption{Results of our linear regression analysis for $L^{'}_{\rm iso}$, $E^{'}_{\rm p}$ and $\Gamma_0$ in the jet co-moving frame by excluding GRB 060218 from our sample , in which $r$ is the Spearman correlation coefficient, $p$ is the chance probability, and $\delta$ is the dispersion of the reported relations.}
\tablecolumns{5}
\tablehead{
\colhead{Relations}&
\colhead{Expressions}&
\colhead{$r$}&
\colhead{$p$}&
\colhead{$\delta^{*}$}
}
\startdata
$L^{\rm '}_{\rm iso}(E^{\rm '}_{\rm iso})$ & $\log L^{\rm '}_{\rm iso, 46}=(1.13\pm 0.13)+(0.67\pm0.14)\times \log E_{\rm iso,50}$ & 0.66 & $<10^{-4}$ & 0.46 \\
$L^{\rm '}_{\rm iso}(E^{\rm '}_{\rm p})$ & $\log L^{\rm '}_{\rm iso, 46}=(1.19\pm 0.08)+(1.18\pm0.13)\times \log (E^{\rm '}_{\rm p}/{\rm keV}$) & 0.85 & $<10^{-4}$   & 0.41\\
$L^{\rm '}_{\rm iso}(\Gamma_0)$ & $\log L^{\rm '}_{\rm iso, 46}=(2.45\pm0.85)-(0.36\pm0.37)\times \log \Gamma_0$ & -0.17 & 0.34 & -- \\
$E^{\rm '}_{\rm iso}(E^{\rm '}_{\rm p})$ & $\log E^{\rm '}_{\rm iso, 50}=(0.46\pm 0.12)+(0.77\pm0.20)\times \log (E^{\rm '}_{\rm p}/{\rm keV})$ & 0.57 & $6.00\times 10^{-4}$ & --\\
$E^{\rm '}_{\rm iso}(\Gamma_0)$ & $\log E^{\rm '}_{\rm iso, 50}=(0.22\pm0.84)+(0.23\pm0.37)\times \log \Gamma_0$ & 0.11 & 0.53  & -- \\
$E^{\rm '}_{\rm p}(\Gamma_0)$ & $\log (E^{\rm '}_{\rm p}/{\rm keV})=(1.47\pm0.59)-(0.48\pm0.26)\times \log \Gamma_0$ & -0.32 & 0.07 & -- \\
\hline
$L^{\rm 'r}_{\rm iso}(E^{\rm '}_{\rm p},\Gamma_0)$ & $\log L^{\rm 'r}_{\rm iso, 46}=(0.64\pm0.49)+(1.23\pm0.14)\times \log (E^{\rm '}_{\rm p}/{\rm keV})+(0.23\pm0.21)\times \log\Gamma_0$ & 0.86 & $<10^{-4}$& 0.27 \\
$E_{\rm p}^{\rm ' r}(L^{\rm '}_{\rm iso},\Gamma_0)$ & $\log (E_{\rm p}^{\rm ' r}/{\rm keV})=(0.02\pm0.35)+(0.59\pm0.07)\times \log L^{\rm '}_{\rm iso, 46}-(0.27\pm0.14)\times \log \Gamma_0$ & 0.87 & $<10^{-4}$ & 0.19 \\
$\Gamma^{\rm r}_0(L^{\rm '}_{\rm iso},E^{\rm '}_{\rm p})$ & $\log \Gamma^{\rm r}_0=(2.15\pm0.19)+(0.17\pm0.15)\times \log L^{\rm '}_{\rm iso, 46}-(0.42\pm0.21)\times \log (E^{\rm '}_{\rm p}/{\rm keV})$ & 0.37 & 0.03  & -- \\
$E^{\rm 'r}_{\rm iso}(E^{\rm '}_{\rm p},\Gamma_0)$ & $\log E^{\rm 'r}_{\rm iso, 50}=-(1.13\pm0.72)+(0.91\pm0.20)\times \log (E^{\rm '}_{\rm p}/{\rm keV})+(0.67\pm0.30)\times \log\Gamma_0$ & 0.66 & $<10^{-4}$ & 0.30 \\
$E_{\rm p}^{\rm ' r}(E_{\rm iso}^{\rm '},\Gamma_0)$ & $\log (E_{\rm p}^{\rm ' r}/{\rm keV})=(1.37\pm0.46)+(0.45\pm0.10)\times \log E_{\rm iso, 50}^{\rm '}-(0.58\pm0.20)\times \log \Gamma_0$ & 0.68 & $<10^{-4}$ & 0.22 \\
$\Gamma^{\rm r}_0(E^{\rm '}_{\rm iso},E^{\rm '}_{\rm p})$ & $\log \Gamma^{\rm r}_0=(2.26\pm0.07)+(0.21\pm0.09)\times \log E^{\rm '}_{\rm iso, 50}-(0.37\pm 0.13)\times \log (E^{\rm '}_{\rm p}$/{\rm keV}) & 0.48 & $4.91\times 10^{-3}$& -- \\
\enddata
\tablenotetext{*}{Reported only for the relation with $p<10^{-4}$.}

\end{deluxetable}


\clearpage

  \begin{figure*}
   \centering
   \label{figure1}
\includegraphics[angle=0,scale=0.23]{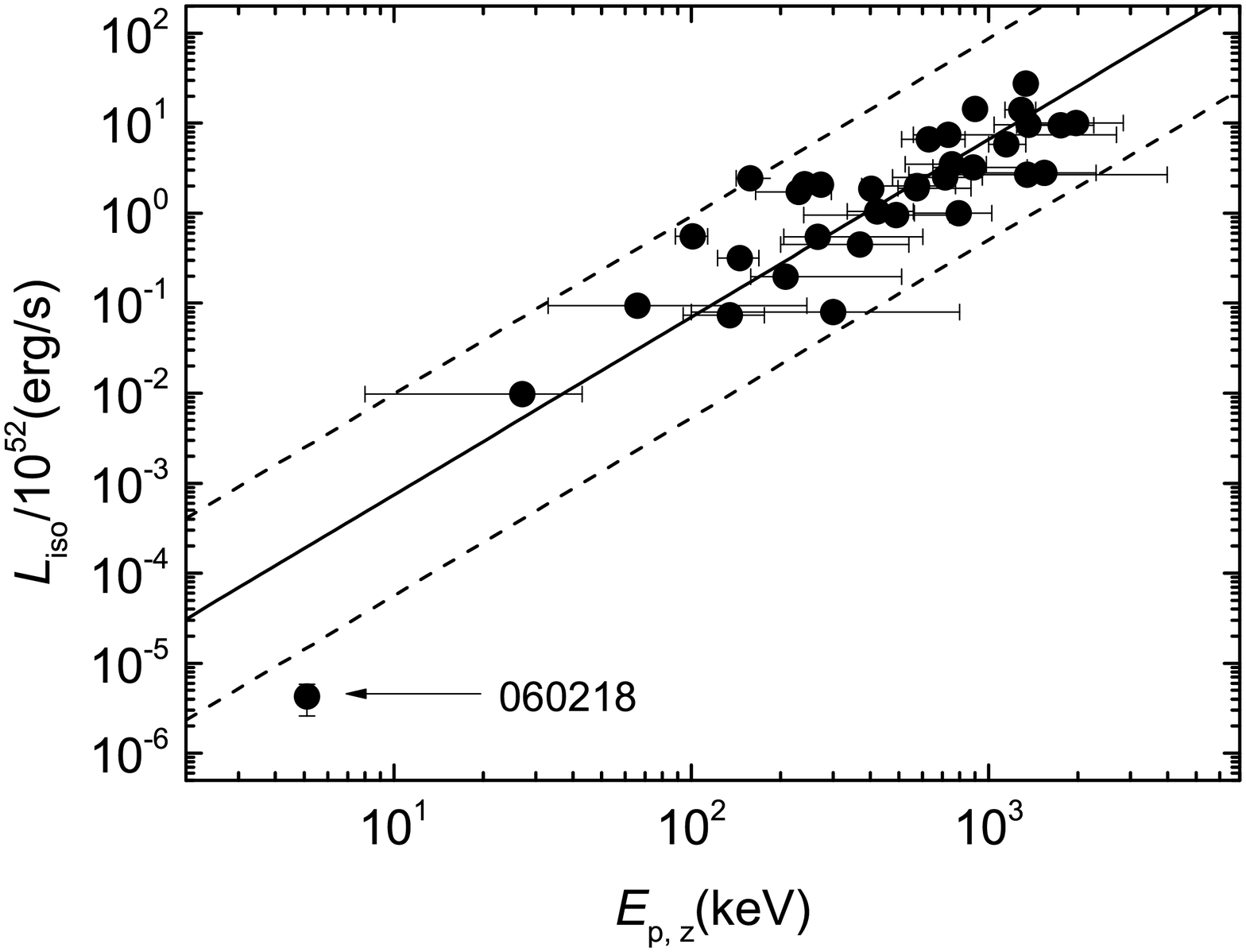}
\includegraphics[angle=0,scale=0.23]{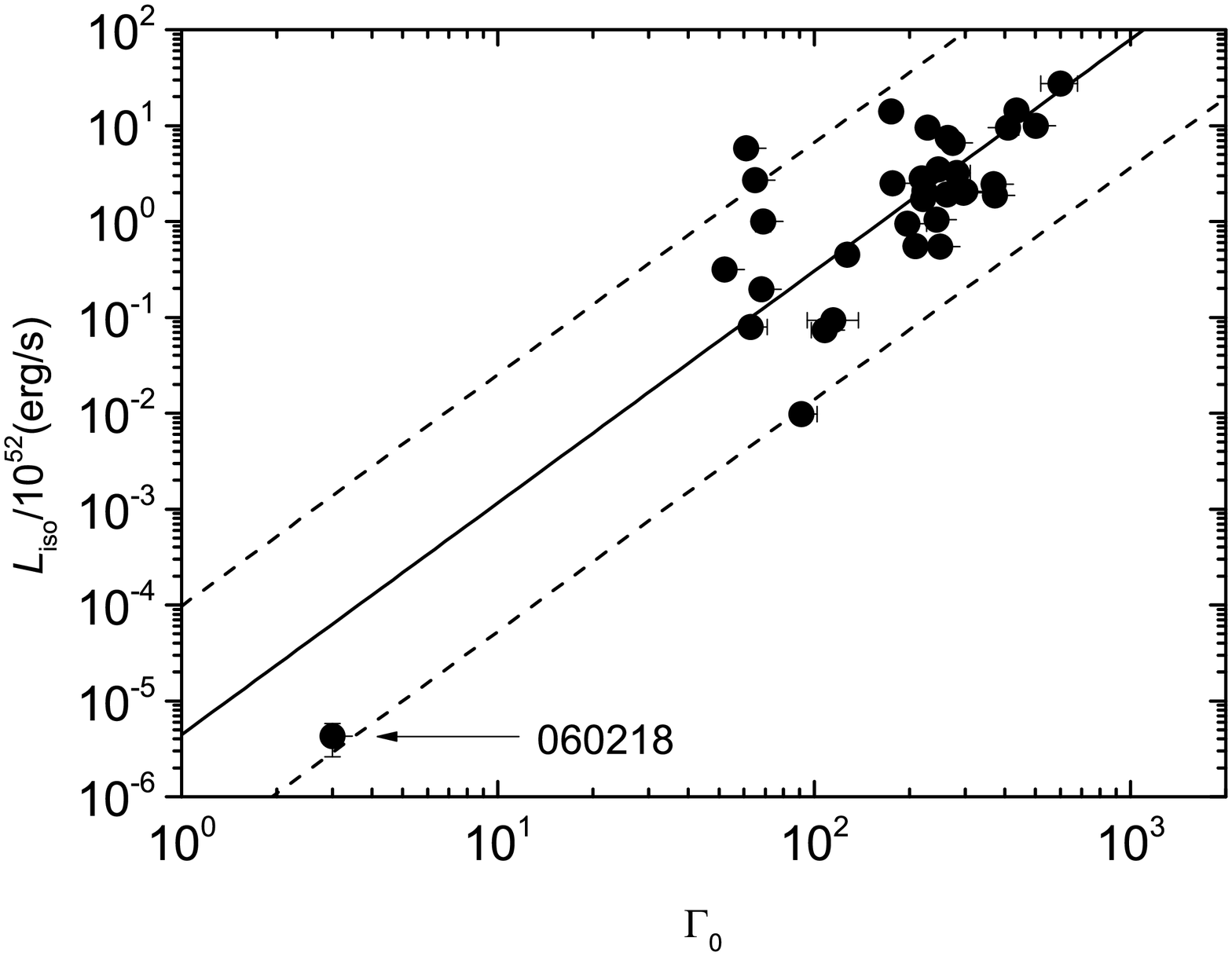}
\includegraphics[angle=0,scale=0.23]{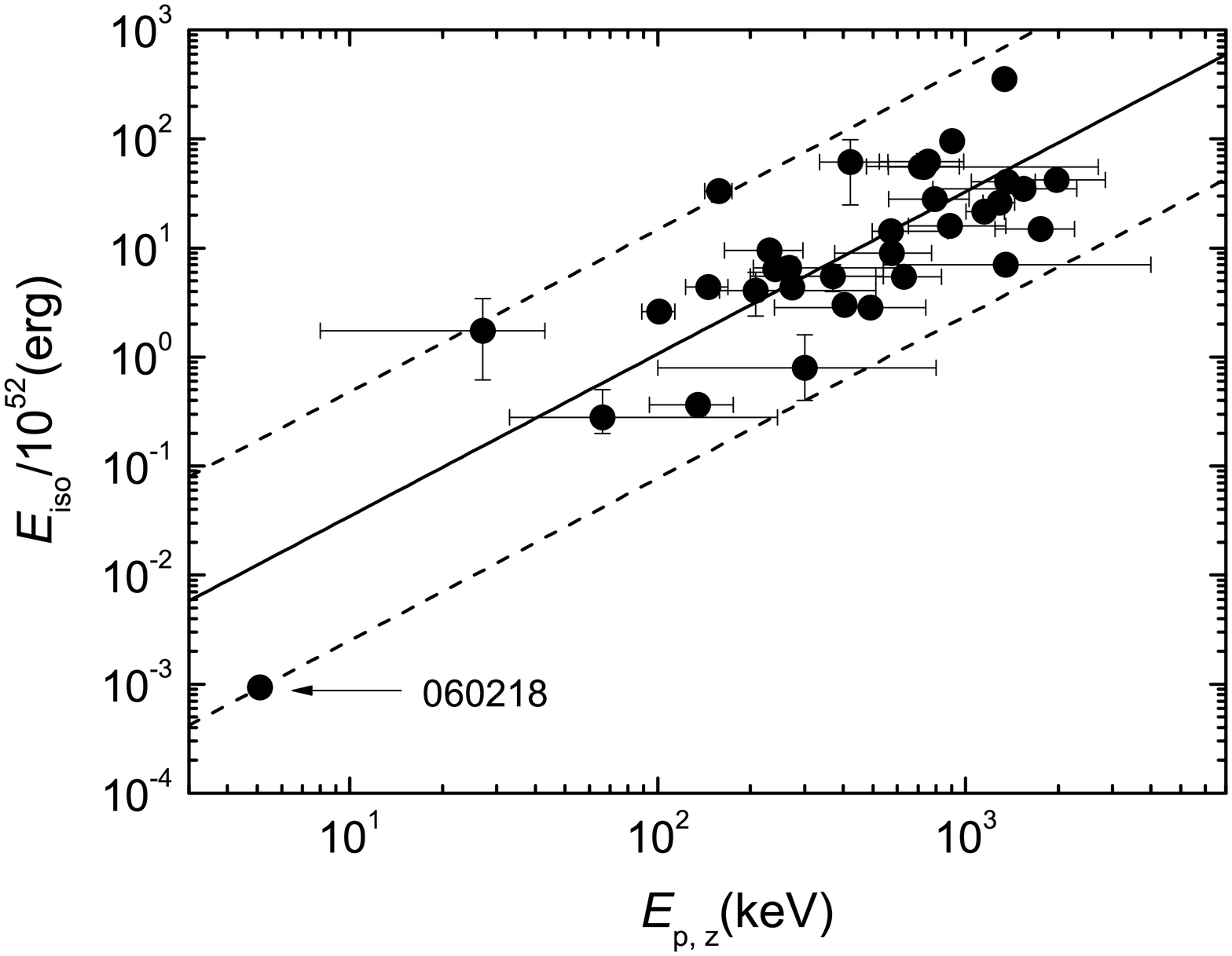}
\includegraphics[angle=0,scale=0.23]{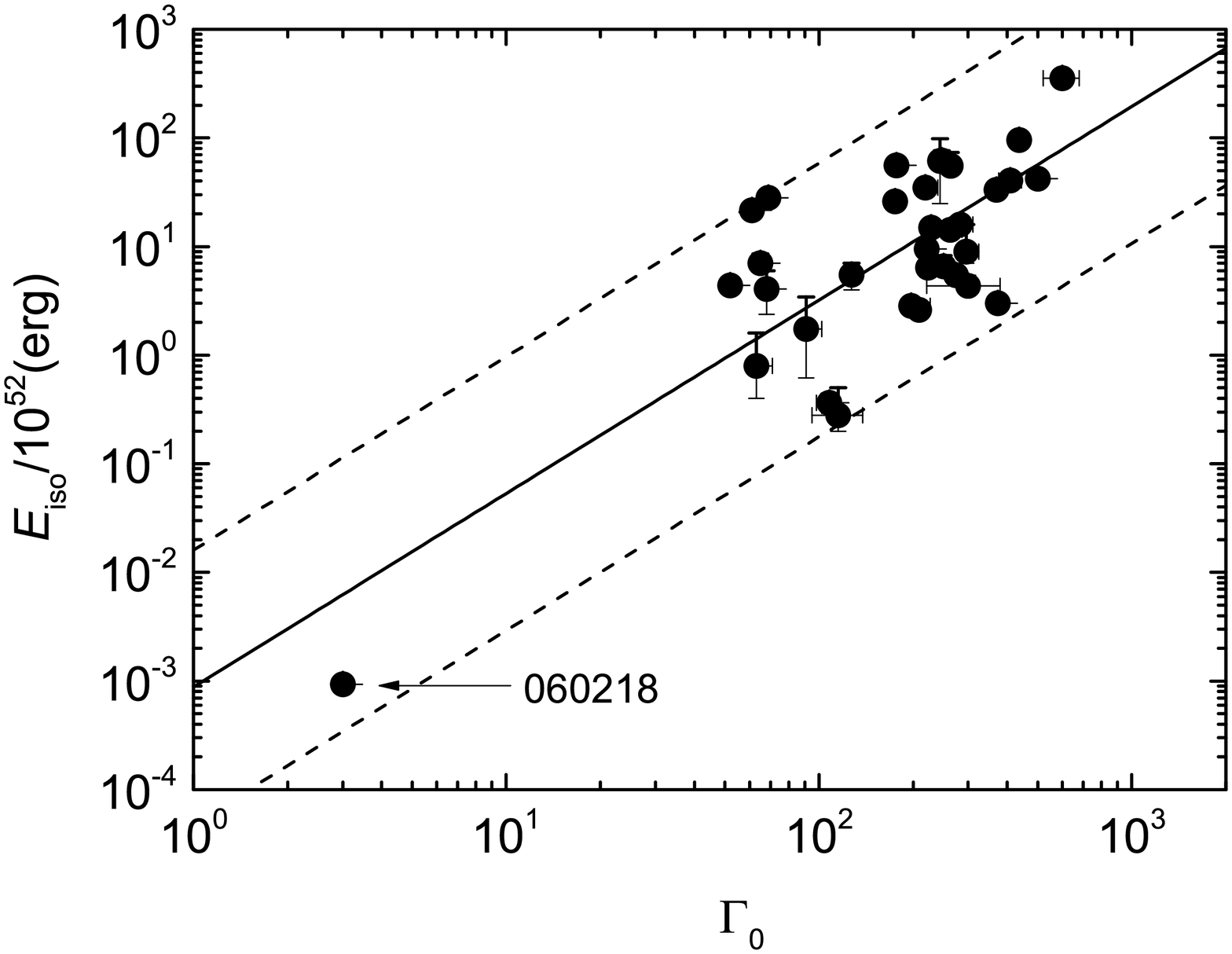}
\includegraphics[angle=0,scale=0.23]{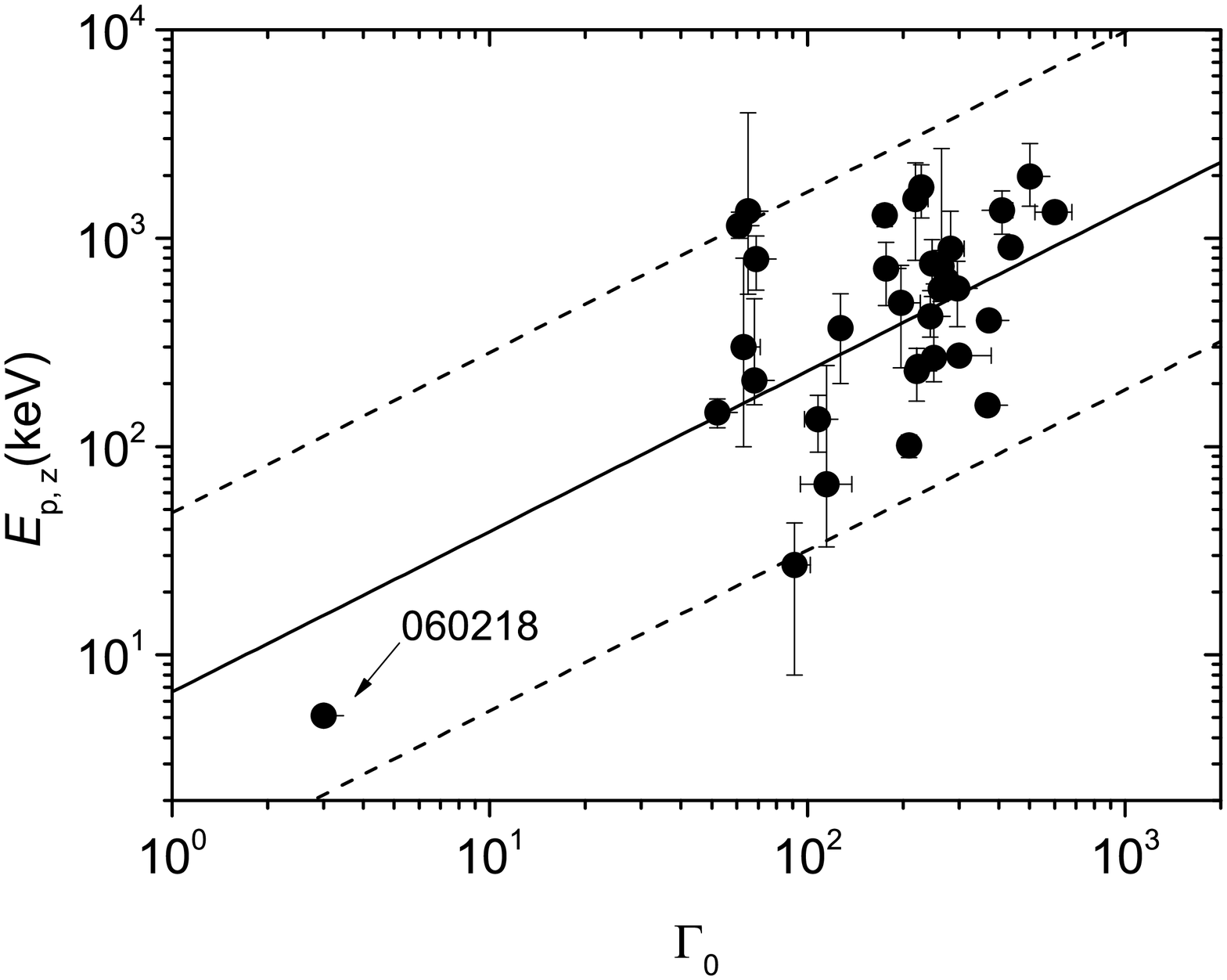}
\caption{Pair correlations among parameter sets \{$L_{\rm p,iso}$, $E_{\rm p,z}$, $\Gamma_0$\} and \{$E_{\rm iso}$, $E_{\rm p,z}$, $\Gamma_0$\}. The best fit lines together with their 2$\sigma$ dispersion regions are shown with solid and dashed lines, respectively. }
  \end{figure*}
\clearpage

  \begin{figure*}
   \centering
\includegraphics[angle=0,scale=0.23]{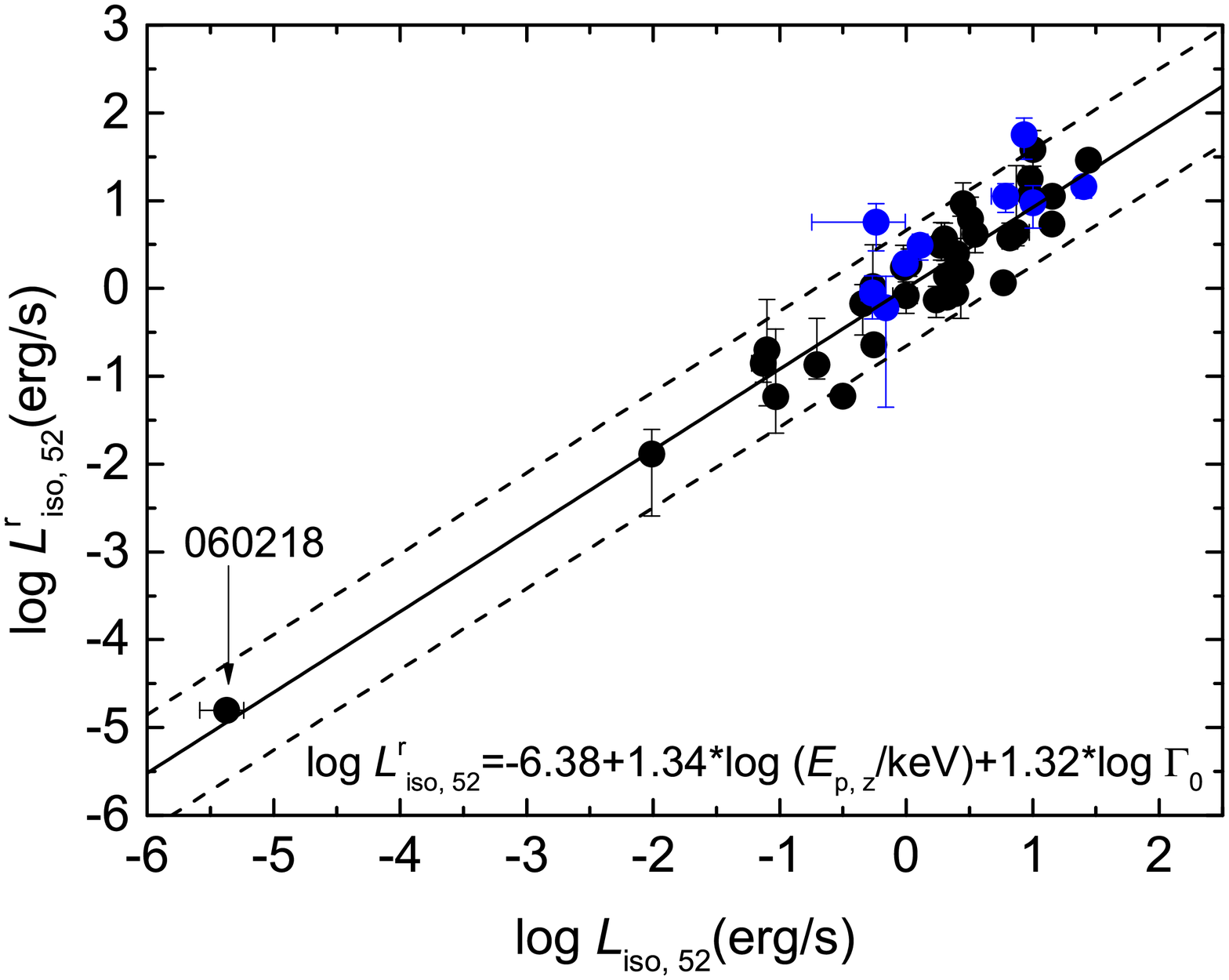}
\includegraphics[angle=0,scale=0.23]{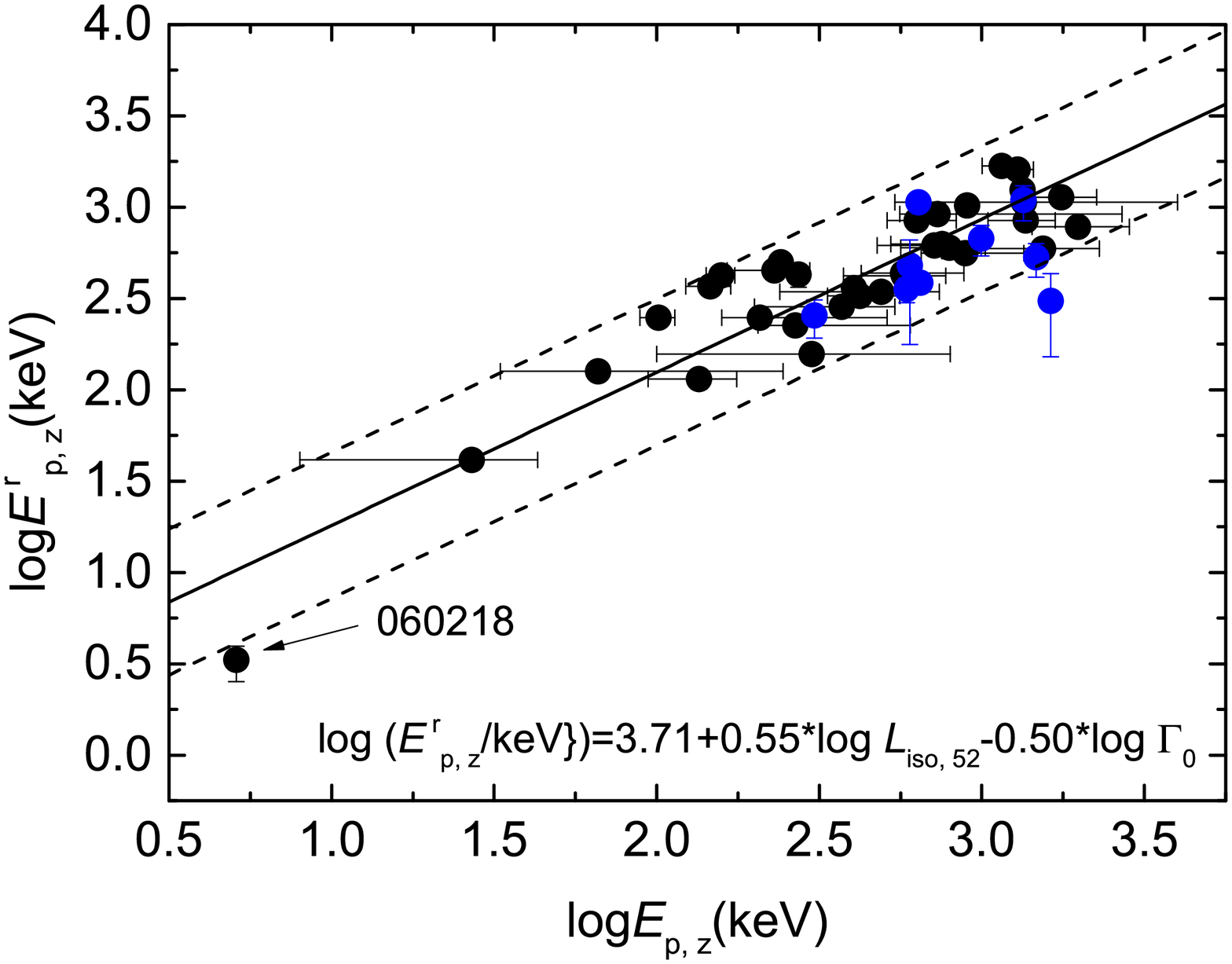}
\includegraphics[angle=0,scale=0.23]{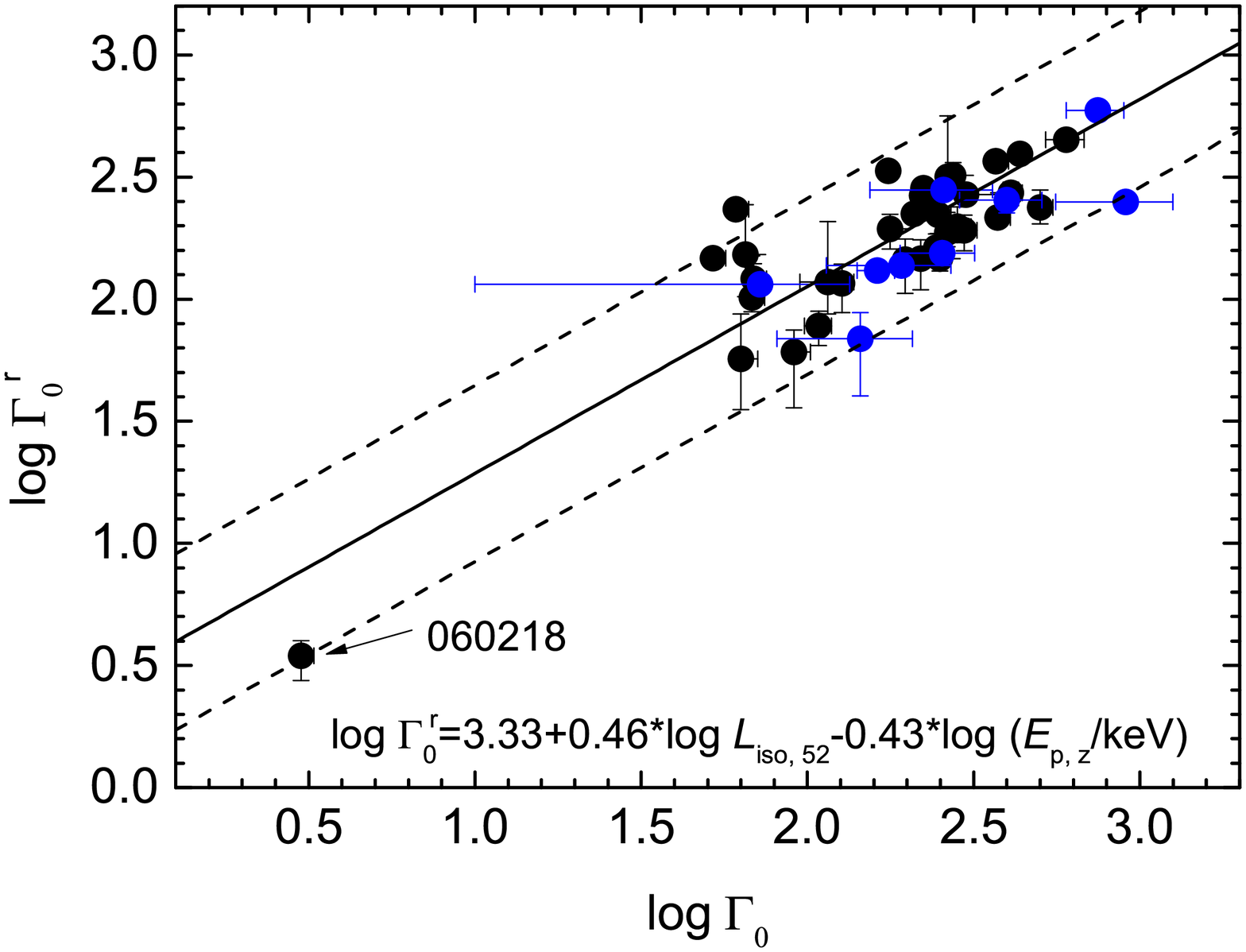}
\includegraphics[angle=0,scale=0.23]{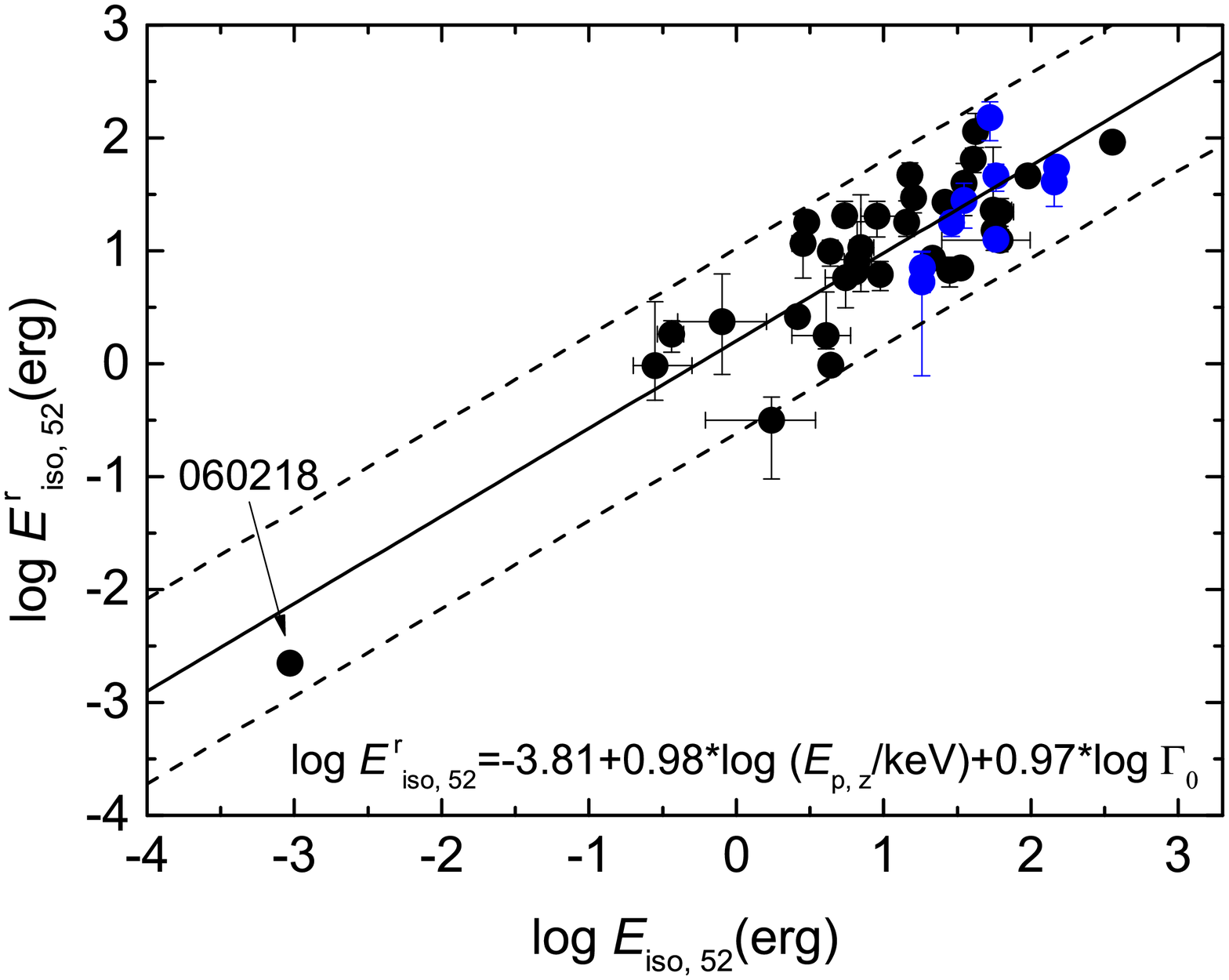}
   \caption{Three-parameter correlations
derived from our multiple regression analysis. The parameters based on the correlations as marked in each panel are compared against the observed parameters. The best fit lines together with their 2$\sigma$ dispersion regions are shown with solid and dashed lines, respectively. For a self-consistency check, 9 GRBs whose $\Gamma_0$ are derived with the opacity argument by using the high energy spectral cutoffs from the joint LAT/GBM observations are also shown (blue dots, taken from Tang et al. 2015). }
   \label{figure2}
  \end{figure*}

  \begin{figure*}
   \centering
\includegraphics[angle=0,scale=0.2]{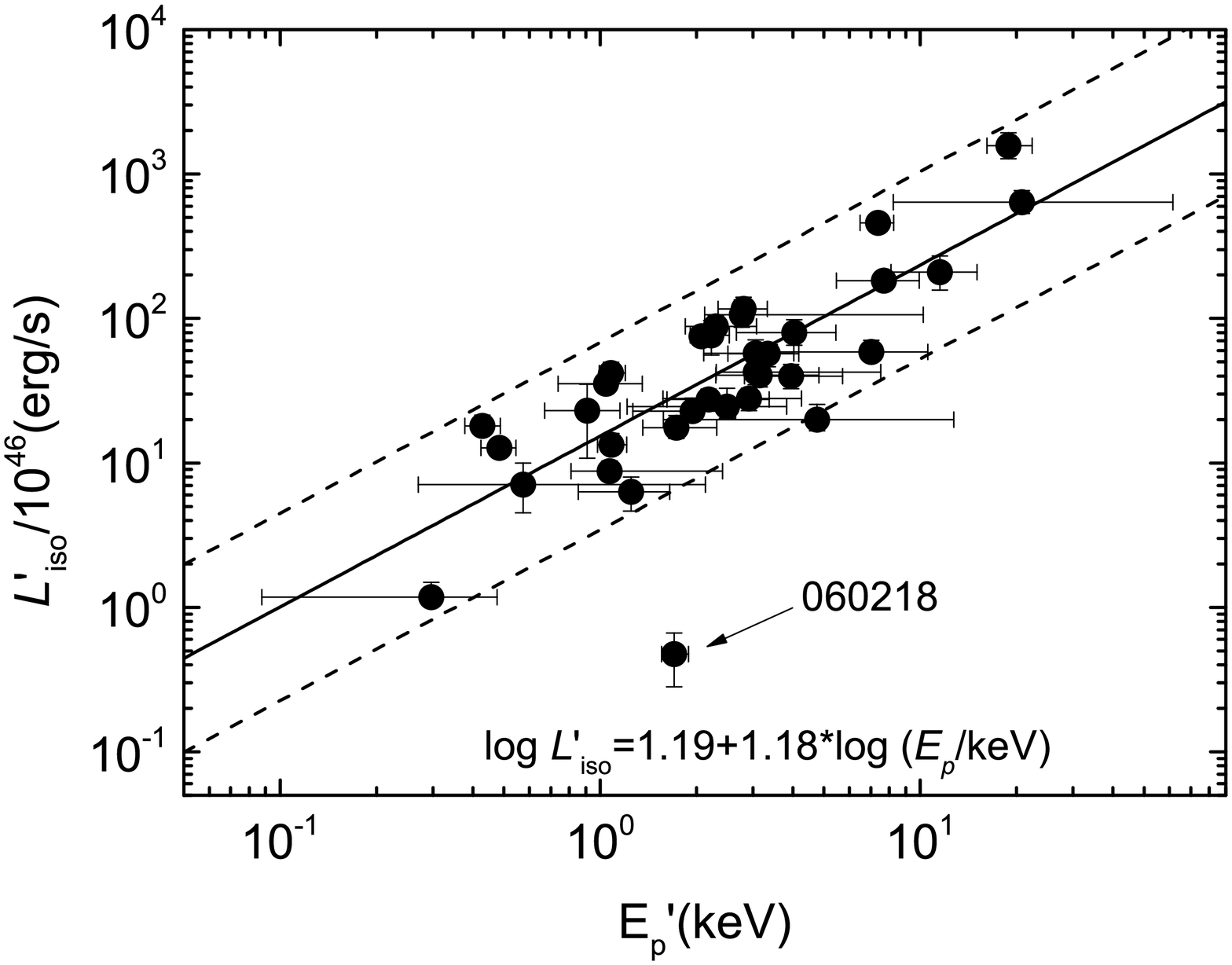}
\includegraphics[angle=0,scale=0.2]{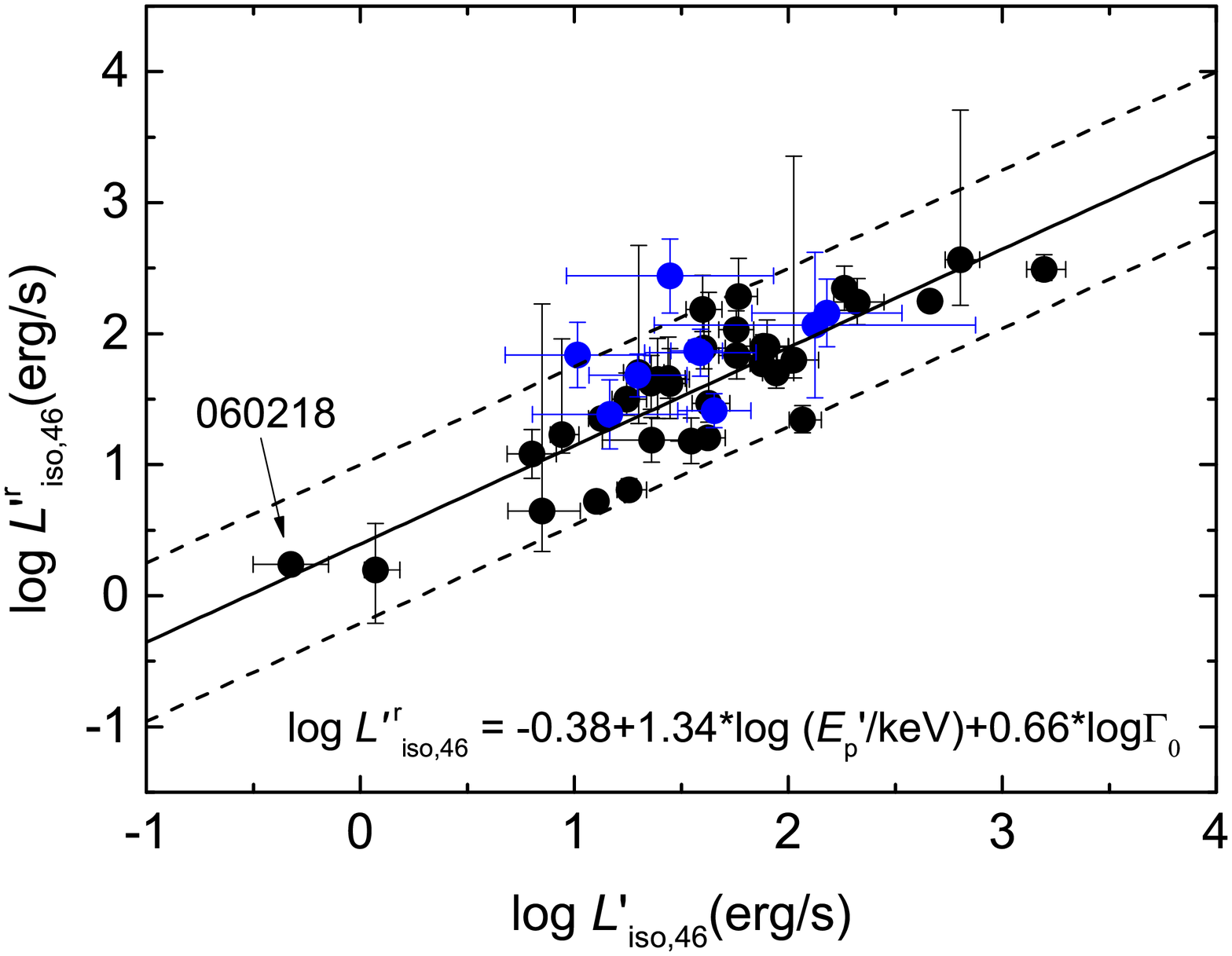}
\includegraphics[angle=0,scale=0.2]{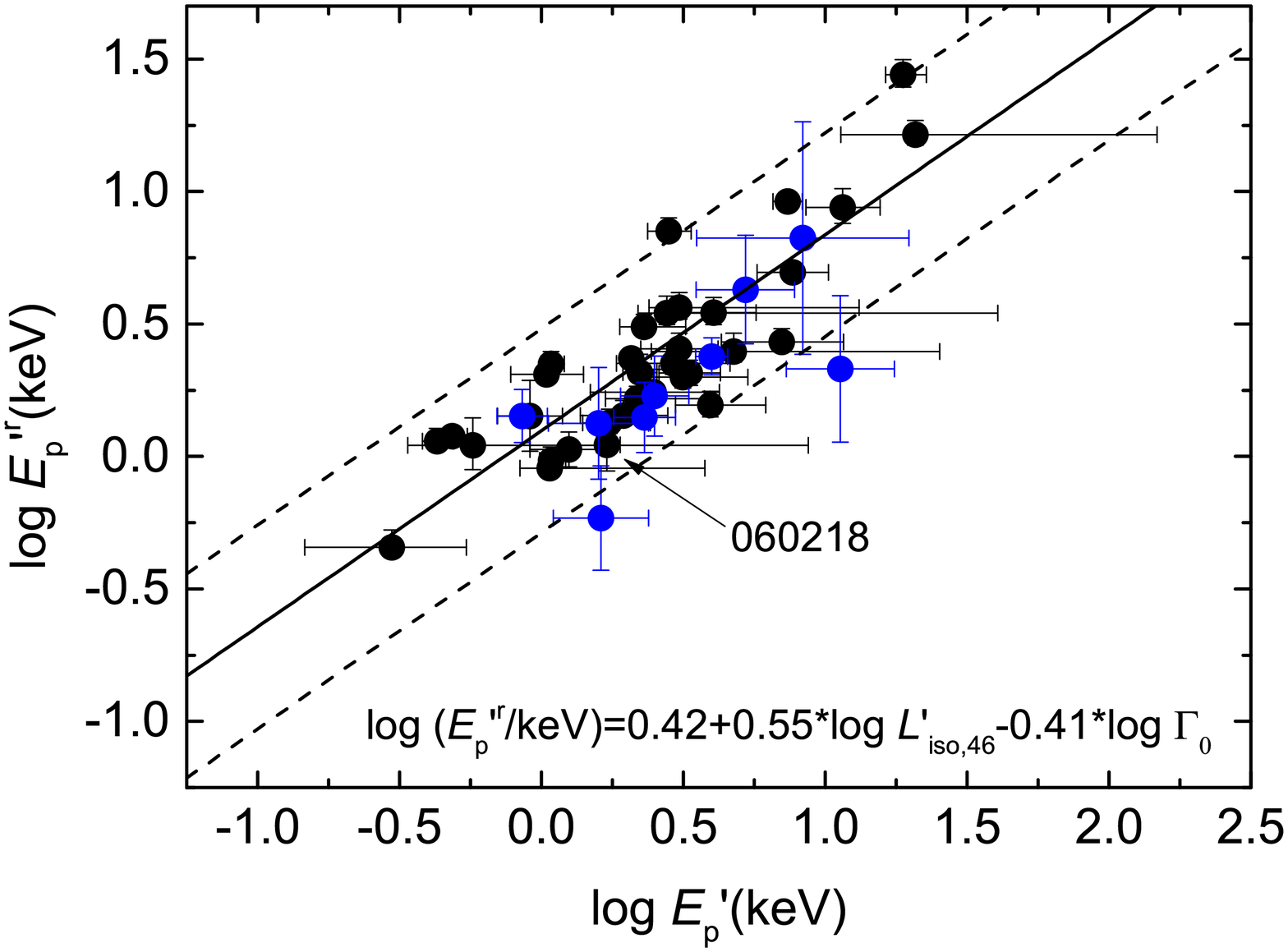}
   \caption{$L^{'}_{\rm iso}-E^{'}_{\rm p}$ and $L^{'}_{\rm iso}-E^{'}_{\rm p}-\Gamma_0$ relations in the jet co-moving frame. The symbol style is the same as Figure 2. An $L^{'}_{\rm iso}-E^{'}_{\rm p}$ relation is found for typical GRBs, i.e., $\log L^{\rm '}_{\rm iso, 46}=(1.19\pm 0.08)+(1.18\pm0.13)\times \log (E^{\rm '}_{\rm p}/{\rm keV}$), but GRB 060218 deviates from this relation at a 3 $\sigma$ confidence level. GRB 060218 shares the same $L^{'}_{\rm iso}-E_p^{'}-\Gamma_0$ relation with typical GRBs.}
   \label{figure3}
  \end{figure*}

\end{document}